# Universal Behavior of the Thermoelectric Figure of Merit, zT, vs. Quality Factor

Evan Witkoske[1*], Xufeng Wang[1], Jesse Maassen[2], and Mark Lundstrom[1]

[1]Purdue University, School of Electrical and Computer Engineering, Purdue University, West Lafayette, IN 47907
[2]Department of Physics and Atmospheric Science, Dalhousie University, Halifax, NS, Canada, B3H 4R2
*Corresponding author at ewitkosk@purdue.edu

**Abstract:** To increase the performance of thermoelectric materials, the electronic parameters in the figure of merit must be improved. In this paper, we use full, numerical band structures and solve the Boltzmann equation in the relaxation time approximation using energy-dependent scattering times informed by first principles simulations. By varying the strength of the electron-phonon coupling or the lattice thermal conductivity, we compute the thermoelectric figure of merit, *zT*, vs. a generalized thermoelectric quality factor. More than a dozen different complex electronic structures are examined. Surprisingly, we find that at a given quality factor, none provides a better figure of merit than that of a material with a simple, parabolic band and acoustic deformation potential scattering. A qualitative argument for this unexpected finding is presented. This apparent universal behavior suggests that even for complex electronic band structures, the thermoelectric figure of merit depends solely on the ratio of electrical to thermal conductivity; the Seebeck coefficient and Lorenz number need not be considered. This observation should simplify the search for promising new materials, but if exceptions to this behavior can be identified, new paths for increasing thermoelectric material performance will open up.
*(Keywords: Thermoelectrics, Figure of Merit zT, Quality factor, B factor)*

## I. Introduction

The performance of a thermoelectric material is directly related to its material figure of merit,

$$zT = \frac{S^2 \sigma T}{\kappa_L + \kappa_e}, \qquad (1)$$

where $S$ is the Seebeck coefficient, $\sigma$ the electrical conductivity, $\kappa_L$ and $\kappa_e$ the lattice and electronic thermal conductivities, and $T$ is the temperature. One way to improve $zT$ is to reduce the lattice thermal conductivity without significantly degrading the electronic properties [1]. Over the past two decades, this strategy has been quite successful [2]–[5]; $\kappa_L$ is approaching the practical lower limit of $\kappa_L = 0.2 \text{ W/m-K}$ identified in [1]. The material figure of merit was $zT \approx 1$ when [1] was written; since then, there have been several reports of $zT > 2$ (e.g. [4], [6], [7]). Device and manufacturing issues must be addressed to turn recent advances in material performance into improved device performance [8], but additional improvements in material performance are also needed. It seems likely that progress in reducing $\kappa_L$ will slow; so further advances in $zT$ will have to come by improving the electronic performance. Several ideas to enhance the electronic performance of thermoelectric materials have been proposed, but success has been elusive.

The relation of $S$, $\sigma$, and $\kappa_e$ to TE properties is well understood within the context of a simple, parabolic band model [9], [10] for which the only thing that matters is the magnitude of the TE quality factor (B-factor), which is proportional to the ratio of the electrical conductivity to the lattice thermal conductivity [11]–[13]. Such analyses suggest that a high $zT$ will not be possible in a simple semiconductor – implying that high performance, if possible at all, will only occur in materials with unusual or complex electronic features [14]. Here we define a "complex" band structure as any band structure other than a simple parabolic band. Because so many factors are involved and because the electronic transport coefficients are so tightly coupled, identifying a complex material with the promise of significantly out-performing a simple parabolic band is hard to do in a clear and convincing way. Over the past two decades, many proposals to enhance the electronic performance of thermoelectrics have been presented [15]–[22]. There is a need for a clear way to compare complex TE materials and engineered structures on a common basis to determine whether there is any electronic structure that can significantly out-perform a simple parabolic energy band.

In this paper, we present simulations that suggest a plot of the peak $zT$ (i.e. the $zT$ at the optimum Fermi level, $\hat{E}_F$) vs. a generalized quality factor at the peak $zT$ is a universal characteristic. We present results for more than a dozen widely different electronic structures and show that none exceeds the performance of a simple, parabolic band material with acoustic deformation potential scattering. A simple argument explains the results and suggests that no complex thermoelectric material will out-perform a material with simple, parabolic energy bands. The result, however, is not fundamental, if materials that substantially exceed the parabolic band limit can be found, they will provide new routes to increasing $zT$.

**II. Approach**

A brief description of the computational techniques employed follows; a more extensive discussion can be found in the supplementary information.

Equation (1) can be re-expressed as

$$zT = \frac{S'^2}{L' + 1/b_L}, \qquad (2)$$

where $S' = S/(k_B/q)$ is the dimensionless Seebeck coefficient, $L' = L/(k_B/q)^2$ the dimensionless Lorenz number, and

$$b_L(E_F) \equiv \frac{\sigma(E_F)T}{\kappa_L}(k_B/q)^2 \qquad (3)$$

is a generalized b-factor; it is closely related (see [23]) to the B-factor discussed by Mahan [9], [11], which is also the "material factor" $\beta$ introduced by Chasmar and Stratton [12]. The important role that the B-factor (also called the quality factor) plays in thermoelectric materials has been discussed by Wang et al. [13]. As noted by Mahan, in the absence of bipolar effects, $zT$ is a function of $B$ (or $b_L$) alone and does not depend independently on the parameters, $S$, $\sigma$ or $\kappa_L$ [9].

The relation between B and $b_L(E_F)$ is simple and is given by eqn. (9) in Sec. V. The key difference is that $b_L(E_F)$ depends on Fermi level and B does not. The second key difference is that $b_L(E_F)$ is defined for any band structure while B is only defined for parabolic energy bands. Because our focus is on complex band structures, we will work with $b_L(E_F)$ in this paper. To calibrate readers, note that for parabolic bands with B = 0.4, we find $b_L(E_F \approx E_C) = 0.18$. For more discussion, see the supplementary information.

The material figure of merit increases without limit as the b-factor increases, but the magnitude of the b-factor is, however, not the whole story. In this paper, we ask the question: "Are there materials or engineered structures that provide *at the same b-factor* a higher *zT* than a material with a simple parabolic band?" Large b-factors will always be necessary, but if the answer to this question is yes, new options for increasing *zT* will open up.

By solving the Boltzmann Transport Equation in the Relaxation Time Approximation, we find the thermoelectric transport parameters as

$$\sigma = \int_{-\infty}^{+\infty} \sigma'(E) dE \qquad (4a)$$

$$S = -\frac{1}{qT} \int_{-\infty}^{+\infty} (E - E_F) \sigma'(E) dE \bigg/ \int_{-\infty}^{+\infty} \sigma'(E) dE \qquad (4b)$$

$$\kappa_0 = \frac{1}{q^2 T} \int_{-\infty}^{+\infty} (E - E_F)^2 \sigma'(E) dE = \kappa_e + T\sigma S^2 \qquad (4c)$$

where the differential conductivity, $\sigma'(E)$, is

$$\sigma'(E) = q^2 \Xi(E) \left( -\partial f_0 / \partial E \right), \qquad (4d)$$

and the transport distribution in the diffusive limit [24],

$$\Xi(E) = \frac{2}{h} \left( M(E)/A \right) \lambda(E), \qquad (4e)$$

is written in Landauer form with $M(E)/A$ being the number of channels per cross-sectional area for conduction and $\lambda(E)$ the mean-free-path (MFP) for backscattering. In (4c), $\kappa_0$ is the electronic thermal conductivity measured under short circuit conditions, and $\kappa_e$ is the same quantity measured under open circuit conditions. See the appendix in [23] for a short derivation of (4e) and [24] for a longer discussion.

In (4e), the mean-free-path for backscattering is defined as [24]

$$\lambda(E) \equiv 2v_x^2(E)\tau_m(E)/v_x^+(E), \tag{5a}$$

where $v_x^2(E)$ is an average over angle of the quantity $v_x^2(\vec{k})$ at energy, $E$. The velocity, $v_x^+(E)$, is the angle-averaged velocity in the +x direction (see [24] for the definitions of these averages). The number of channels at energy, $E$, is [24], [25]

$$M(E)/A = hv_x^+(E)D(E)/4, \tag{5b}$$

where $D(E)$ is the density-of-states per unit volume including a factor of two for spin.

The treatment of electron scattering is an important consideration in any TE calculation. The use of an energy-independent momentum relaxation time, $\tau_m$ can lead to errors [23], [26]. A better assumption when isotropic electron-phonon scattering dominates is that the scattering rate follows the density-of-states [1],

$$\frac{1}{\tau(E)} = \frac{1}{\tau_m(E)} \propto K_{el-ph} D(E), \tag{5c}$$

where $K_{el-ph}$ describes the electron-phonon coupling, $\tau$ is the scattering time, and $\tau_m$ is the momentum relaxation time. Equation (5c) commonly describes phonon scattering in non-polar semiconductors [23], [27]. As discussed [26], it also seems to describe some TE materials and appears to be a much better approximation than the assumption of a constant scattering time. Rigorous treatments of electron scattering are available to provide material-specific scattering times [28]–[32]. Finally, we note that when plotting $zT$ vs. $b_L$, $K_{el-ph}$ is not needed because $S'$ and $L'$ only depend on the energy dependence of the scattering rate – not on its magnitude.

### III. Results: Parabolic Energy Bands

Equations (4) and (5) can be solved numerically given a full, numerical description of the energy band (see the supplementary material of [23], [26]). Equations (2) and (3) can also be solved analytically for simple parabolic band structures (the results are presented in the appendix of [25]). For parabolic energy bands with power law scattering,

$$\lambda(E) = \lambda_0 \left[(E - E_C)/k_B T\right]^r, \tag{6}$$

where *r* is a characteristic exponent. For acoustic deformation potential (ADP) scattering in 3D parabolic bands, $r = 0$; the MFP is independent of energy. For ionized impurity (II) scattering, $r = 2$. By assuming a temperature, effective mass, MFP ($\lambda_0$), characteristic exponent, $r$, and lattice thermal conductivity, $\kappa_L$, one can sweep the Fermi level and produce a plot of $zT(E_F)$ vs. $b_L(E_F)$. Results are shown in Fig. 1 for several different values of $\kappa_L$.

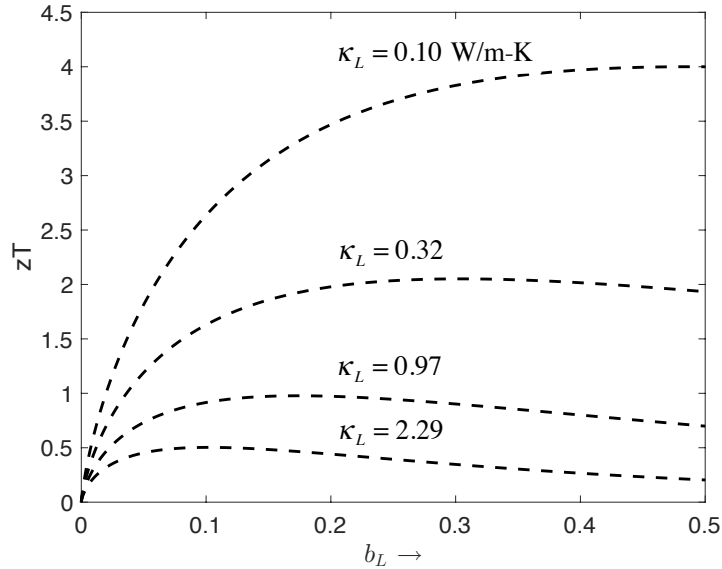

Fig. 1. Material figure of merit, $zT$ vs. $b_L$ for $\kappa_L = 0.10$, 0.32, 0.97, and 2.29 W/m-K, which were selected to produce peak zT's of 4, 2, 1, and 0.5 using a n-type Bi$_2$Te$_3$ parabolic conduction band with $m^* = 1.56 m_0$ and an energy-independent ($r = 0$) MFP of $\lambda_0 = 25$ nm. As the Fermi level increases, $b_L$ increases because $\sigma(E_F)$ increases.

For each $\kappa_L$ in Fig. 1, there is a maximum in $zT$ as the Fermi level is swept. By plotting $zT$ vs. $b_L$ at the peak $zT$ where $E_F = \hat{E}_F$, the results in Fig. 2 are obtained. Each point on the $zT(\hat{E}_F)$ vs. $b_L(\hat{E}_F)$ characteristic is the value of $zT$ and $b_L$ at the Fermi level that maximizes $zT$. The $zT(\hat{E}_F)$ vs. $b_L(\hat{E}_F)$ characteristic is independent of how $b_L$ is varied (i.e. by varying $\kappa_L$ as in Fig. 1, the MFP parameter $\lambda_0$, or the effective mass). The characteristic is also independent of temperature. (We should note, however, that these calculations do not include bipolar conduction; when they are considered, the characteristic is temperature dependent and sensitive to the ratio of mean-free-paths and effective masses.) Figure 2 compares the $zT(\hat{E}_F)$ vs. $b_L(\hat{E}_F)$ characteristic for 3D, parabolic bands with ADP scattering ($r=0$) to that for II scattering ($r=2$). Also shown is the single energy channel case [1].

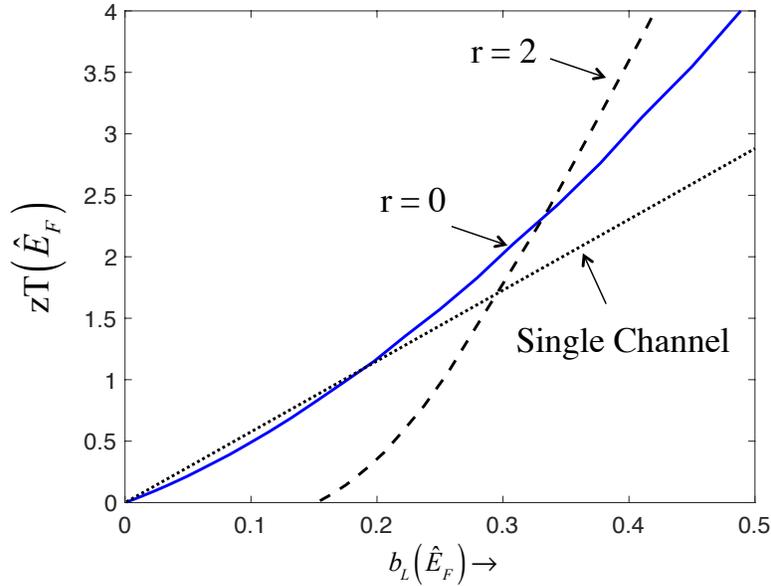

Fig. 2.  Peak material figures of merit, $zT(\hat{E}_F)$ vs. $b_L(\hat{E}_F)$, the value of $b_L$ at the Fermi level that maximizes $zT$. Solid line: Parabolic bands with $r=0$ in eqn. (6b). Dashed line: Parabolic bands with $r=2$. Dotted line: Single energy channel.

Finally, the distinction between $b_L(E_F)$ and $b_L(\hat{E}_F)$ should be kept in mind. Figure 1 is a plot of $zT(E_F)$ vs. $b_L(E_F)$. The Fermi level that maximizes $zT$ is $E_F = \hat{E}_F$, Figure 2 is a plot of $zT(\hat{E}_F)$ vs. $b_L(\hat{E}_F)$.

For ADP scattering ($r=0$) $|S'(\hat{E}_F)|$ increases with $b_L(\hat{E}_F)$ because as $b_L$ increases, the Fermi level at peak $zT$ drops below the band edge (see Figs. 1a and 4 of [23]). The Lorenz number is close to its non-degenerate limit of two and saturates at two for large $b_L(\hat{E}_F)$.

For II scattering ($r = 2$), the peak Fermi level for small b-factors lies deep in the conduction band where $|S|$ and therefore $zT$ approach zero [33]. For $b_L > 0.35$, the Fermi level at the peak $zT$ moves below the band edge, and II scattering produces a high Lorenz number approaching four, which is detrimental because $L'$ is in the denominator of (2). For large b-factors however, the very large $S'^2$ for II scattering causes this case to out-perform the ADP scattering case, as shown in Fig. 2. The II scattering case shows that it is possible to exceed the parabolic band/ADP scattering limit with the right type of energy-dependent scattering. In practice, II scattering often occurs in addition to electron-phonon scattering, which lowers the overall MFP and reduces $zT$. The point here is that with the right energy dependence to the transport function (which is proportional to the product of the number of channels and the MFP, eqn. (4e)), then the parabolic band / ADP limit can be exceeded.

Figure 2 also compares the peak $zT(\hat{E}_F)$ vs. $b_L(\hat{E}_F)$ characteristics for parabolic bands to the single energy channel case,

$$M(E)/A = M_0 \delta(E - E_C), \tag{7}$$

where $E_C$ is the energy of the channel. The single energy channel case displays n-type conduction when $E_F < E_C$. In contrast to the parabolic band case, the Fermi level at the peak $zT$ is independent of $b_L$, i.e. $(E_F - E_C)/k_B T = -2.4 = S'$. For a single energy channel, $L' = 0$ [1], so equation (2) gives $zT = (S')^2 b_L = 5.75 b_L$. As shown in Fig. 2, a single energy channel provides little benefit over a parabolic band when $zT < 1$ and under-performs a parabolic band when $zT > 1$. These results support the conclusion of [34], [35] that a delta-function transport distribution is not the best for thermoelectric performance, but the magnitude of the b-factor must be considered as well.

Analytical results provide reference points for comparison to the full, numerical solutions of the thermoelectric equations that are considered next. As we examine the influence of complex band structures on the $zT(\hat{E}_F)$ vs. $b_L(\hat{E}_F)$ characteristic, the parabolic band with ADP scattering case will serve as our reference because ADP scattering is thought to dominate in many TE materials [10]. Note that for ADP scattering in parabolic bands ($r = 0$ in eqn. (6)), the scattering rate follows the parabolic band density-of-states.

**IV. Results: Complex Energy Bands**

Thermoelectric materials typically have complex band structures with multiple, anisotropic bands or pockets. In this paper, we will refer to any band structure that is more complicated than a single, parabolic band as "complex." Recent work shows, for example, that the Lorenz number can be significantly different from the value computed for a parabolic band [26]. Higher $zT's$ might be possible with materials that offer a higher Seebeck coefficient or a lower Lorenz number (e.g. [5], [15]–[19], [21], [36]). To examine this possibility, we present full, numerical simulations analogous to the analytical calculations discussed above. The numerical methods used are described in the supplementary information of [23] and [26]. The key input is a band structure from density functional theory (DFT) simulations.

Figure 3 compares the computed $zT(\hat{E}_F)$ vs. $b_L(\hat{E}_F)$ characteristics using density of states (DOS) scattering as described by eqn. (5c) for nine complex band structures, p-$Bi_2Te_3$, n-$Bi_2Te_3$, p-SnSe, n-SnSe, p-$Sb_2Te_3$, n-$Bi_{0.85}Sb_{0.15}$, p-$Bi_{0.85}Sb_{0.15}$, an n-$Bi_2Te_3$ quintuple layer, a p-$Bi_2Te_3$ quintuple layer, p-$Mg_3Sb_2$, and p-GeTe, to that of a simple, parabolic energy band with ADP scattering(r = 0). The $zT(\hat{E}_F)$ vs. $b_L(\hat{E}_F)$ characteristics shown in Fig. 3 are insensitive to how the b-factor is varied. The b-factor can be changed by varying $\kappa_L$ in equation (1) or by varying the electron-phonon coupling parameter in (5c), which varies $\sigma$ in equation (1), but just as we found for the analytical calculations, the same $zT(\hat{E}_F)$ vs. $b_L(\hat{E}_F)$ characteristic is obtained. The numerical calculations also show that the characteristic is relatively insensitive to temperature. (We remind the reader, however, that bipolar effects are not considered.)

The cases shown in Fig. 3 explore a broad (though not exhaustive) range of complex band structures. What stands out in Fig. 3 is the fact that the results of numerical calculations for a variety of materials with complex band structures are remarkably similar to the analytical calculations assuming a parabolic band with ADP scattering. The p-$Bi_2Te_3$ quintuple layer is a 2D material with a band structure that is thought to be advantageous for thermoelectrics [37], but its performance is no better than that of a material with a simple parabolic energy band. These results suggest that only the magnitude of the b-factor matters. For a given b-factor, all thermoelectric materials seem to provide nearly the same peak $zT$. As discussed next, however, these results assume a particular treatment of scattering.

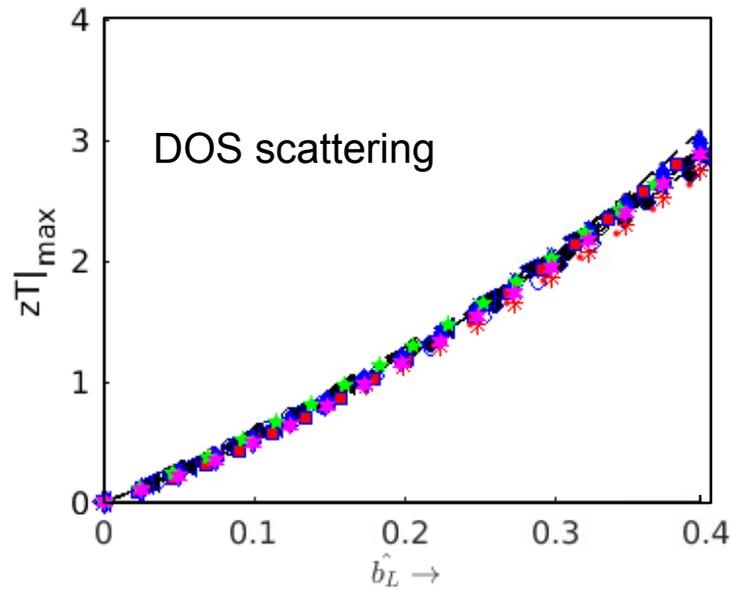

Fig. 3    Material figure of merit, peak zT vs. $\hat{b}_L$ at 300K, for several different complex thermoelectric materials assuming a scattering rate proportional to the density of states. Open circles are for p-$Bi_2Te_3$ and filled circles for n-$Bi_2Te_3$. Open squares are for p-SnSe and filled squares for n-SnSe. Filled triangles are for p-$Sb_2Te_3$. Asterisks are for n-$Bi_{0.85}Sb_{0.15}$ and x-marks are for p-$Bi_{0.85}Sb_{0.15}$. Stars are for an n-$Bi_2Te_3$ quintuple layer, and diamonds are for p-$Bi_2Te_3$ quintuple layer. Pentagons are p-$Mg_3Sb_2$ and red dots are p-GeTe. The dashed line is the parabolic band reference assuming $r = 0$.

**V. Discussion**

Two key factors in the calculations presented in Fig. 3 are the band structure and electron scattering. The band structures were computed by DFT simulation. The results presented in Fig. 3 assumed that the electron-phonon scattering rate follows the total density-of-states according to (5c) with the assumption that the intra-valley/band coupling strength is equal to the inter-valley/band coupling strength. Rigorous treatments of electron-phonon scattering suggest that it is a good assumption for Si [23] and for SnSe [26]. If we were to suppress inter-valley/band scattering, the $zT(\hat{E}_F)$ vs. $b_L(\hat{E}_F)$ characteristic would not change. What would change is that for a given electron-phonon coupling strength, the conductivity with inter-valley scattering would be lower than the conductivity without inter-valley scattering. The b-factor, and therefore, $zT$, would change respectively.

To examine the sensitivity of the results to the treatment of carrier scattering, we repeated the calculations with the assumption of a constant mean-free-path. The results are shown in Fig. 4. Although the constant mean-free-path assumption produces more spread about the parabolic band reference, the results shown in Figs. 3 and 4 show that the $zT(\hat{E}_F)$ vs. $b_L(\hat{E}_F)$ characteristic is relatively insensitive to the details of band structure and carrier scattering.

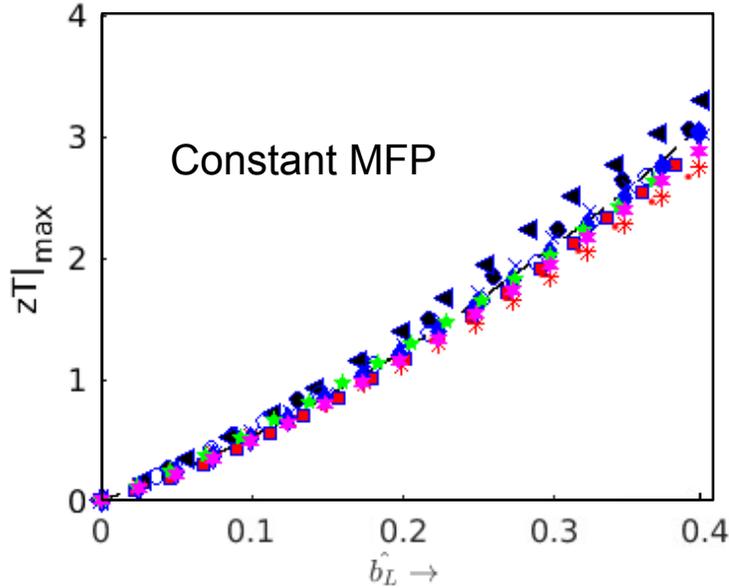

Fig. 4  Material figure of merit, peak zT vs. $\hat{b}_L$ at 300K, for several different complex thermoelectric materials assuming a constant MFP for scattering. Open circles are for p-Bi$_2$Te$_3$ and filled circles for n-Bi$_2$Te$_3$. Open squares are for p-SnSe and filled

squares for n-SnSe. Filled triangles are for p-Sb$_2$Te$_3$. Asterisks are for n-Bi$_{0.85}$Sb$_{0.15}$ and x-marks are for p-Bi$_{0.85}$Sb$_{0.15}$. Stars are for an n-Bi$_2$Te$_3$ quintuple layer, and diamonds are for p-Bi$_2$Te$_3$ quintuple layer. Pentagons are p-Mg$_3$Sb$_2$ and red dots are p-GeTe. The dashed line is the parabolic band reference assuming $r = 0$.

Equation (1) provides a qualitative explanation for the insensitivity of the $zT(\hat{E}_F)$ vs. $b_L(\hat{E}_F)$ characteristic to band structure and scattering physics. A transport function that increases $S$ also increases $L$, which limits the benefit to $zT$. Conversely, a transport function that reduces $L$ also reduces $S$, again limiting the benefit to $zT$. This conjecture is confirmed by Fig. 5, which is a plot of $S(\hat{E}_F)$ and $L'(\hat{E}_F)$ vs. $b_L(\hat{E}_F)$ for p-Bi$_2$Te$_3$, p-SnSe, and p-Sb$_2$Te$_3$ (DOS scattering was assumed for these calculations). Figure 5 shows that $S(\hat{E}_F)$ and $L'(\hat{E}_F)$ depend on band structure and can be quite different than for a parabolic band. In the range of interest, ($zT(\hat{E}_F) > 2 \Rightarrow b_L(\hat{E}_F) > 0.25$), the three materials shown all display Lorenz numbers that are well below that of the parabolic band reference [26], but this advantage is offset by the fact that they all display lower Seebeck coefficients as well. The trade-off between $S'$ and $L'$ makes it difficult to enhance $zT$ in comparison to a simple, parabolic band.

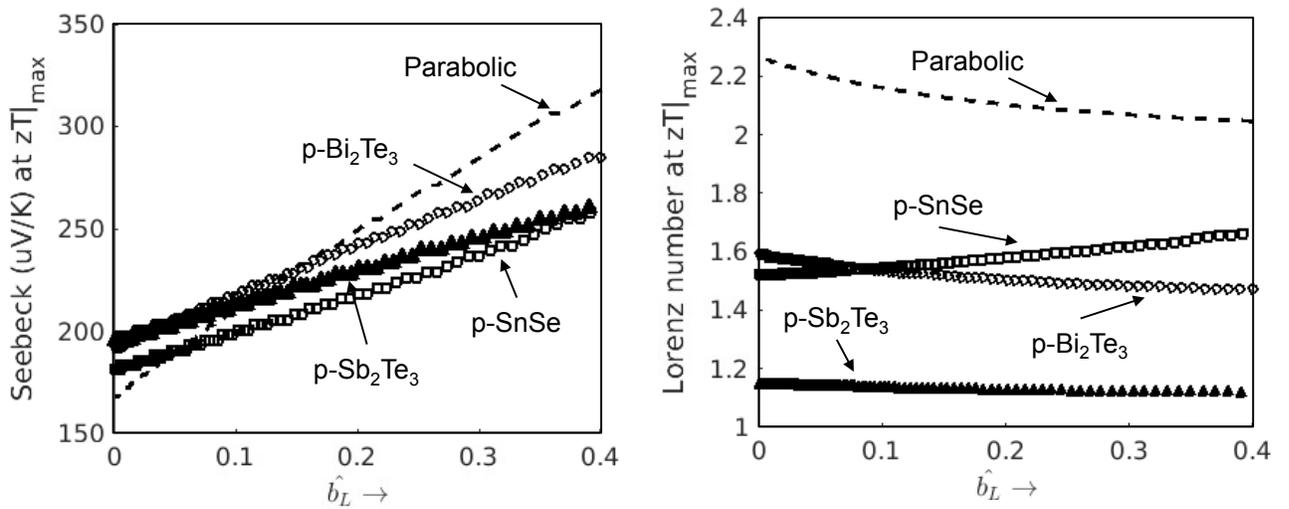

Fig. 5    Seebeck coefficient and Lorenz number at peak zT vs. $\hat{b}_L$ at 300 K for several different complex thermoelectric materials assuming a scattering rate

proportional to the density of states. Open circles are for p-Bi$_2$Te$_3$. Open squares are for p-SnSe and closed triangles are for p-Sb$_2$Te$_3$. The dashed line is the parabolic band reference assuming $r = 0$.

The results of this paper suggest a different, possibly simpler way to assess the performance of a thermoelectric material. First, one should check to see how it compares at a given b-factor to a parabolic band with ADP scattering. Second, one should assess its potential to achieve a large b-factor. This involves assessing the lattice thermal conductivity and assessing the electrical conductivity, which can be done as follows.

From eqns. (4a), (4d), and (4e), we find

$$\begin{aligned}\sigma &= \frac{2q^2}{h}\int \left[M(E)/A\right]\left[\lambda(E)\right]\left(-\partial f_0/\partial E\right)dE = q^2 \int \frac{v_x^2(E)\left(-\partial f_0/\partial E\right)}{K_{el-ph}}dE \\ &\approx \frac{q^2}{K_{el-ph}}\int v_x^2(E)\left(-\partial f_0/\partial E\right)dE = \frac{q^2 \langle v_x^2(E)\rangle}{K_{el-ph}}\end{aligned}, \qquad (8)$$

where we have assumed that $K_{el-ph}$ varies slowly with energy across the Fermi window. For electrical conductivity, eqn. (8) shows only two things matter; that the material has a low electron-phonon coupling parameter and a high velocity squared in the direction of transport. It does not matter whether the high velocity comes from a single valley with a light effective mass, from an anisotropic valley with a light effective mass in the direction of transport, or from multiple valleys with light effective masses in the direction of transport. Of course, all this assumes that the scattering rate follows the density-of-states. Detailed calculations suggest that this is the case, but this question should be examined in more detail. In this approach, assessing the electrical potential of a material is simplified because one only needs to estimate the electron-phonon coupling parameter, $K_{el-ph}$, and calculate $\langle v_x^2(E)\rangle$ directly from the band structure. This approach should find use in assessing the potential of a given material, and it might also find application in high throughput computational searches [7], [38-41].

Finally, there has been some interest in defining a general quality factor that is independent of band structure [42]. The generalized b-factor given by (3) applies to any band structure, but it includes a dependence on the Fermi level, which is not included in the traditional definition [9], [11]–[13]. For parabolic bands,

$$b_L(E_F) \equiv \frac{\sigma(E_F)T}{\kappa_L}(k_B/q)^2 = B\mathcal{F}_{1/2}\left[(E_F - E_C)/k_B T\right], \qquad (9)$$

where $B$ is the traditional quality factor [9], [11]–[13], and $\mathcal{F}_{1/2}[\eta]$ is the Fermi-Dirac integral as defined by Blakemore [43]. A general quality factor could be defined as [42]

$$b'_L(E_F) \equiv \frac{\sigma(E_F)T(k_B/q)^2}{\kappa_L \mathcal{F}_{1/2}\left[(E_F - E_C)/k_B T\right]}. \qquad (10)$$

The general quality factor can be evaluated for any band structure and is expected to be relatively insensitive to the location of the Fermi level. For parabolic energy bands, $b'_L$ reduces to the Fermi level independent *B*-factor.

**VI. Summary**

A variety of electronic structures and complex thermoelectric materials were examined in this paper, and all were shown to produce nearly the same $zT(\hat{E}_F)$ vs. $b_L(\hat{E}_F)$ characteristic as that given by a simple, parabolic band model. The material figure of merit, *zT*, increases without limit as the quality factor increases, but the results of this study suggest that at a given b-factor, there is an upper limit to *zT*. The inherent trade-offs between thermoelectric material parameters explain the apparent universal behavior that was found, but this result is not fundamental. Just as the Wiedemann-Franz Law is not a law of nature, but rather, a rule of thumb that is only rarely violated [44], the same may be true of the universal behavior discovered here. Searches for materials/structures that exceed the parabolic band limit should be conducted, so that new pathways to higher thermoelectric performance can be identified. If no such materials are found, then searches need only focus on materials with large generalized b-factors, i.e. with high electrical conductivity and low lattice thermal conductivity.

*Acknowledgement* – This work was partially supported by the Defense Advanced Research Projects Agency (Award No. HR0011-15-2-0037). J. Maassen would like to acknowledge support from NSERC (Discovery Grant RGPIN-2016-04881).

*Data Availability* – The raw and processed data required to reproduce these findings are available free online to download from [https://nanohub.org/groups/needs/lantrap].

# Supplementary Material for:

# Universal Behavior of the Thermoelectric Figure of Merit, zT, vs. Quality Factor

Evan Witkoske[1*], Xufeng Wang[1], Jesse Maassen[2], and Mark Lundstrom[1]

[1]Purdue University, School of Electrical and Computer Engineering, Purdue University, West Lafayette, IN 47907
[2]Department of Physics and Atmospheric Science, Dalhousie University, Halifax, NS, Canada, B3H 4R2
*Corresponding author at ewitkosk@purdue.edu

## 1) Analytical solutions for thermoelectric parameters

**Parabolic energy bands:**

For parabolic energy bands in 1D, 2D, or 3D (*d* = 1, 2, or 3) and with power law scattering, we find from the results in the appendix of [25] that

$$\sigma_{1D} = \frac{2q^2}{h}\lambda_0 \Gamma(r+1)\mathcal{F}_{r-1}(\eta_F) \qquad \text{(S-m)} \qquad \text{(A1)}$$

$$\sigma_{2D} = \frac{2q^2}{h}\lambda_0\left(\frac{\sqrt{2m^*k_BT}}{\pi\hbar}\right)\Gamma(r+3/2)\mathcal{F}_{r-1/2}(\eta_F) \qquad \text{(S)} \qquad \text{(A2)}$$

$$\sigma_{3D} = \frac{2q^2}{h}\lambda_0\left(\frac{m^*k_BT}{2\pi\hbar^2}\right)\Gamma(r+2)\mathcal{F}_r(\eta_F) \qquad \text{(S/m)} \qquad \text{(A3)}$$

$$S_{d=1,2,3} = -\left(\frac{k_B}{q}\right)\left(\frac{(r+(d+1)/2)\mathcal{F}_{r+(d-1)/2}(\eta_F)}{\mathcal{F}_{r+(d-3)/2}(\eta_F)} - \eta_F\right) \quad \text{(V/K)} \quad \text{(A4)}$$

$$L'_{d=1,2,3} = L_{d=1,2,3}/(k_B/q)^2 \quad \text{(A5a)}$$

$$L'_{d=1,2,3} = \frac{\Gamma(r+(d+3)/2)}{\Gamma(r+(d+1)/2)}\left[\begin{array}{c}(r+(d+3)/2)\dfrac{\mathcal{F}_{r+(d+1)/2}(\eta_F)}{\mathcal{F}_{r+(d-3)/2}(\eta_F)} \\ -(r+(d+1)/2)\left(\dfrac{\mathcal{F}_{r+(d-1)/2}(\eta_F)}{\mathcal{F}_{r+(d-3)/2}(\eta_F)}\right)^2\end{array}\right], \quad \text{(A6)}$$

where

$$\eta_F = (E_F - E_C)/k_B T \quad \text{(A7)}$$

is the dimensionless Fermi energy (chemical potential) and $\mathcal{F}_j(\eta_F)$ is the Fermi-Dirac integral of order $j$ written in the Blakemore form [42]

$$\mathcal{F}_j(\eta_F) = \frac{1}{\Gamma(j+1)}\int_0^\infty \frac{\eta^j d\eta}{1+e^{\eta-\eta_F}}. \quad \text{(A8)}$$

It should be noted that the radius of the nanowire and thickness of the quantum well, do not appear in (A1) and (A2).

For ADP scattering in 3D, $r=0$, in 2D, $r=1/2$, and 1D, $r=1$, so from (A4) and (A5) we find

$$S_{d=1,2,3} = -\left(\frac{k_B}{q}\right)\left(\frac{2\mathcal{F}_1(\eta_F)}{\mathcal{F}_0(\eta_F)} - \eta_F\right) \quad \text{(A9)}$$

and

$$L_{d=1,2,3} = 2\left(\frac{k_B}{q}\right)^2\left\{3\frac{\mathcal{F}_2(\eta_F)}{\mathcal{F}_0(\eta_F)} - \frac{2\mathcal{F}_1^2(\eta_F)}{\mathcal{F}_0^2(\eta_F)}\right\}. \quad \text{(A10)}$$

For ADP scattering in parabolic bands, the Seebeck coefficient and Lorenz numbers are identical in all dimensions.

**Single energy channel**

Analytical solutions are easy to obtain for the single energy channel case,

$$M(E)/A = M_0 \delta(E - E_C),  \tag{A11}$$

where $E_C$ is the energy of the channel. We find

$$\sigma_\delta = \frac{2q^2}{h} \lambda_0 M_0 \left( -\frac{\partial f_0}{\partial E} \bigg|_{E=E_C} \right) = \frac{2q^2}{h} \lambda_0 \langle M \rangle,  \tag{A12}$$

$$S_\delta = \left( -\frac{k_B}{q} \right) \frac{E_C - E_F}{k_B T},  \tag{A13}$$

and

$$L_\delta = 0.  \tag{A14}$$

The figure of merit for the single energy case is readily shown to be

$$zT = \frac{S^2 \sigma T}{\kappa_L} = \left( \frac{k_B}{q} \right)^2 \left( \frac{E_C - E_F}{k_B T} \right)^2 \frac{2q^2 \lambda_0 M_0 T}{h \times \kappa_L} \left( -\frac{\partial f_0}{\partial E} \bigg|_{E=E_C} \right)  \tag{A15}$$

At the maximum $zT$, $|E_C - E_F|/k_B T = 2.4$ and $\partial f_0/\partial E|_{E=E_C} = 0.076/k_B T$, we find

$$zT\big|_{\max} = 0.88 \left( \frac{k_B}{h} \right) \left( \frac{\lambda_0 M_0}{\kappa_L} \right)  \tag{A16}$$

which is essentially eqn. (21) of [1] in a different notation.

Finally, we compute the peak $zT$ vs. b-factor at the peak characteristic for a single energy channel. Because the Lorenz number for a single energy channel is zero, (2) gives

$$zT = (S')^2 b_L  \tag{A17}$$

At the maximum $zT$, $|E_C - E_F|/k_B T = 2.4$, so $S' = -2.4$, and (A17) gives

$$zT = 5.76 b_L,  \tag{A18}$$

which is the equation of the dotted straight line in Fig. 2.

## 2) Observation of "Double-branch" behavior in zT|max vs. b-factor for certain band structures

Some curious behaviors were observed when investigating hypothetical band structures using the effective mass approach. An example of such a peculiar band structure is shown in Figure S1 (left). It consists of two valence bands with energy offsets of 0.2 eV. The highest valence band (Band #1) is isotropic, and the lower band (Band #2) is anisotropic with the same Density of States (DOS) effective mass, but 5 times the Distribution of Modes (DOM) effective mass in transport direction as Band #1. In other words, Band #2 has the same amount of states but much higher velocity—this is advantageous for obtaining higher TE performance. The resulting DOS and DOM of this band structure are shown in Figure S1 (right).

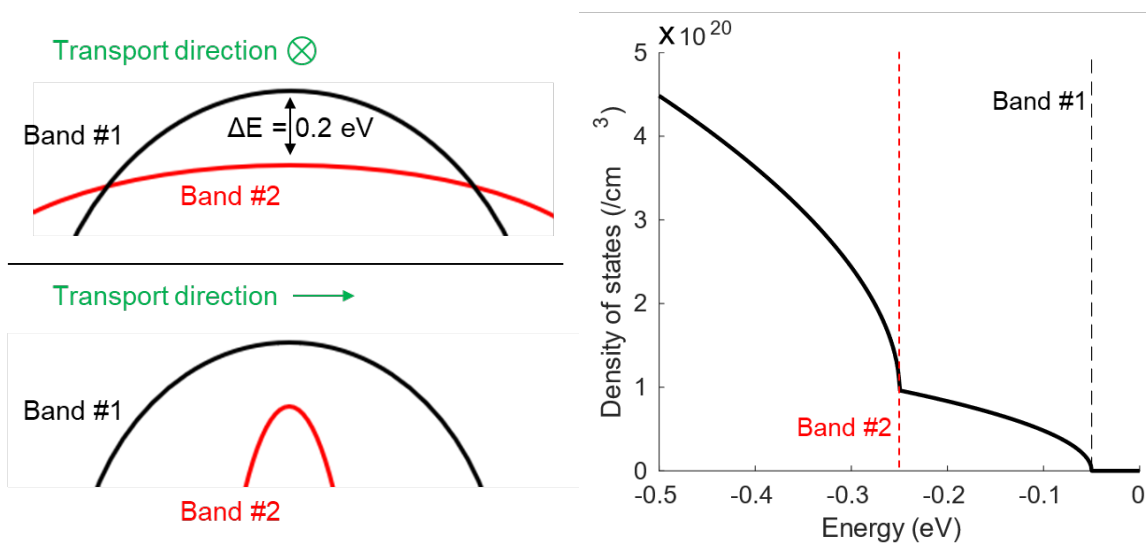

**Figure S1. (Left)** The two-band toy model is used. The highest valence band (black) is 0.2 eV higher than Band #2 (red). Along the transport direction, Band #2 has high velocity. Both bands have the same DOS. **(Right)** The DOS of the two-band model. The edges of Band #1 and Band #2 are marked with black and red dash lines respectively.

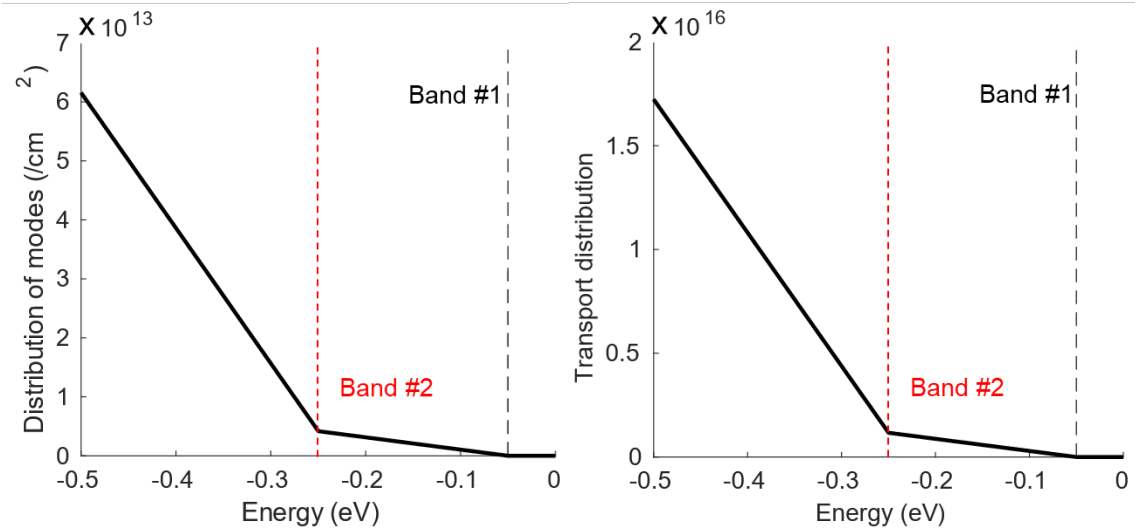

**Figure S2. (Left)** The DOM of the two-band model. **(Right)** The transport distribution of the two-band model. The edges of Band #1 and Band #2 are marked with black and red dash lines respectively.

The peculiarity arises from the optimal location of the Fermi level under a given lattice thermal conductivity, $\kappa_L$. The two bands are essentially in a tight competition with each other for having the maximum zT. The lower band, Band #2, has the advantage of significantly increased DOM, but it also faces the issue that, if the Fermi level is close enough to take advantage of the increased DOM, a significant amount of current will flow on both sides of the Fermi level, due to the presence of Band #1. This is essentially a bipolar effect that decreases the Seebeck coefficient and increases the electronic thermal conductivity—both are undesired and counters the increased electrical conductivity obtained from the increased DOM. This therefore results in a competition between Band #1 and Band #2 for the optimal location of the Fermi level.

This competition is illustrated in Figure S2. Figure S2 (left) shows the situation under a high $\kappa_L$. With a high $\kappa_L$, the electronic thermal conductivity is insignificant, and the optimal location of the Fermi level is decided solely by the power factor. In this case, Band #2 with its high electrical conductivity shows a higher zT than Band #1. However, if $\kappa_L$ is decreased, the benefit of having a lower electronic thermal conductivity quickly catches up for Band #1, and as shown in Figure S2 (right), the maximum zT shifts to favor Band #1.

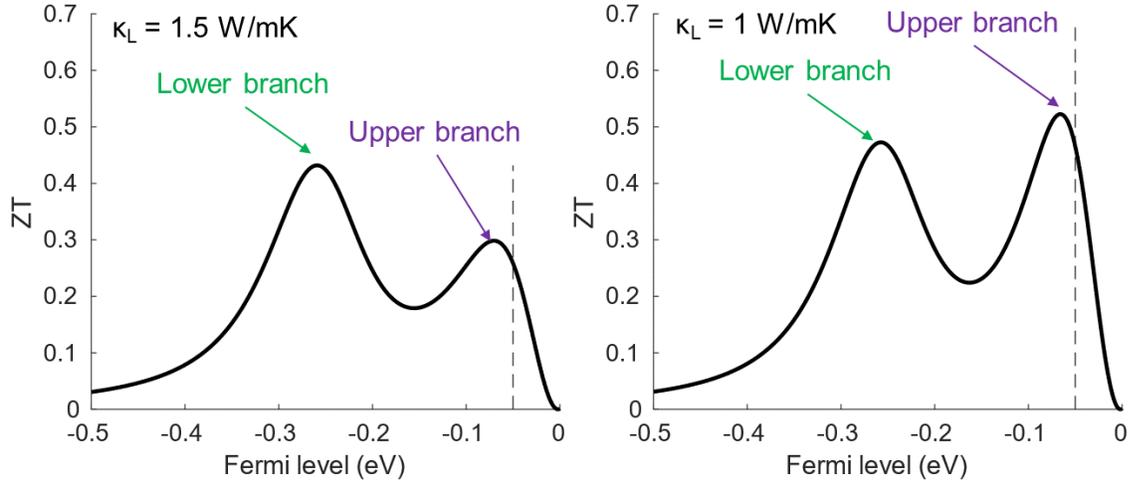

**Figure S3. (Left)** zT vs. Fermi level under high lattice thermal conductivity. **(Right)** zT vs. Fermi level under a moderate lattice thermal conductivity showing the optimal location of the Fermi level for maximum zT shifts. The location of the highest valence band is marked with black dash line.

Because the optimal location of the Fermi level for maximum zT "jumped" from Band #2 to Band #1, it shows up in the zT$|_{max}$ vs. $b_L$ curve as a "snapback" as shown in Figure S4. Since $b_L$ is a ratio between electrical conductivity and lattice thermal conductivity, when the optical location of Fermi level "jumped" from Band #2 to Band #1, the electrical conductivity decreases, causing $b_L$ to decrease. This "snapback" feature forms two distinct branches of the zT$|_{max}$ vs. $b_L$ curve.

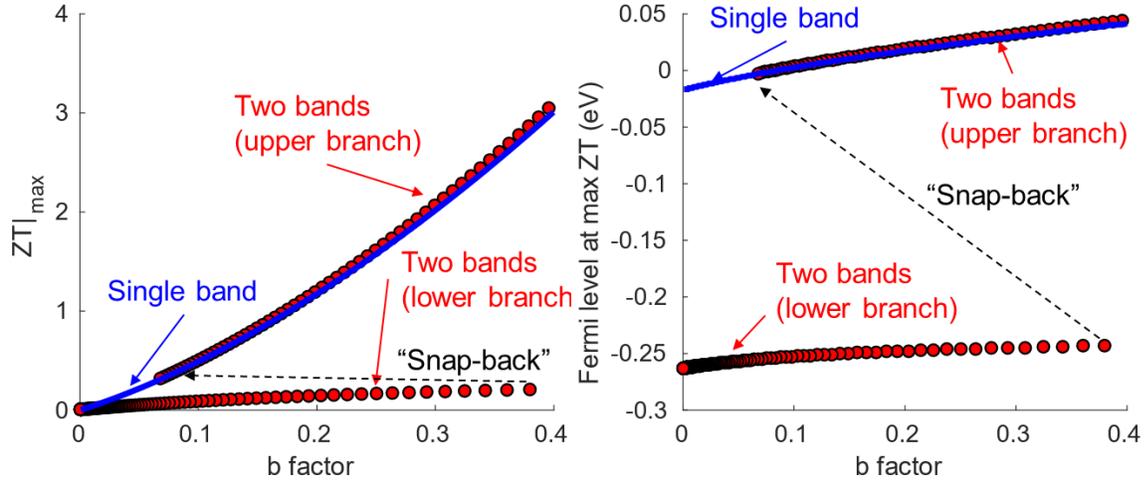

**Figure S4**. Comparison between the example two-band structure and a single parabolic band for **(Left)** zT|$_{max}$ vs. b$_L$ and **(Right)** optimal location of the Fermi level vs. b$_L$.

The location of this "snap-back" does not depend on whether the zT|$_{max}$ vs. b$_L$ curve is obtained via a sweep of $\kappa_L$ or $\sigma$ (i.e. by adjusting the electron-phonon coupling strength). Recall from earlier discussions that the thermoelectric Figure of Merit (FOM) can be defined as follows

$$zT = \frac{S^2 \sigma T}{\kappa_e + \kappa_L} = \frac{S^2/(k_B/q)^2}{\dfrac{\kappa_e}{\sigma T (k_B/q)^2} + \dfrac{\kappa_L}{\sigma T (k_B/q)^2}} = \frac{S'^2}{L' + 1/b_L} \qquad \text{(A18)}$$

with,

$$b_L = \frac{\sigma T}{\kappa_L}\left(\frac{k_B}{q}\right)^2. \qquad \text{(A19)}$$

We will show that it is equivalent to vary $b_L$ through adjusting $\kappa_L$ or $\sigma$. To increase $\sigma$ by a factor of N, one can scale the electron-phonon coupling parameter by a factor of 1/N. The resulting $S$ is unaffected by this scaling, and $\kappa_e$ is scaled by a factor of N. Equation (A18) becomes

$$zT = \frac{S^2 \sigma T}{\kappa_e + \kappa_L} = \frac{S^2/(k_B/q)^2}{\dfrac{N\kappa_e}{N\sigma T (k_B/q)^2} + \dfrac{\kappa_L}{N\sigma T (k_B/q)^2}} = \frac{S'^2}{L' + 1/Nb_L}. \qquad \text{(A20)}$$

Therefore, this has the same effect as scaling $\kappa_L$ by a factor of 1/N.

## 3) A comparison of B, $b_L(\hat{E}_F)$, and $zT(\hat{E}_F)$

The conventional B-factor is defined for parabolic bands and does not include a Fermi level dependence. For parabolic bands, the relation between this and $b_L(E_F)$ is given by eqn. (9) in the paper.

$$b_L(E_F) \equiv \frac{\sigma(E_F)T}{\kappa_L}(k_B/q)^2 = B\mathcal{F}_{1/2}\left[(E_F - E_C)/k_B T\right], \quad (A21)$$

is given by eqn. (9) in the paper.

Our motivation for using a generalized quality factor is that it allows us to deal directly with complex band structures without extracting effective masses and allows inclusion of more general scattering models (e.g. beyond power law). It is also, in principle directly measurable because it depends only on the measured electrical and lattice thermal conductivities.

The quantity $b'_L$,

$$b'_L(E_F) \equiv \frac{\sigma(E_F)T(k_B/q)^2}{\kappa_L \mathcal{F}_{1/2}\left[(E_F - E_C)/k_B T\right]}. \quad (A22)$$

defined in eqn. (10) of the paper is more similar to B, but it is hard to make general quantitative statements because $b'_L$ depends on the specifics of the complex band structure.

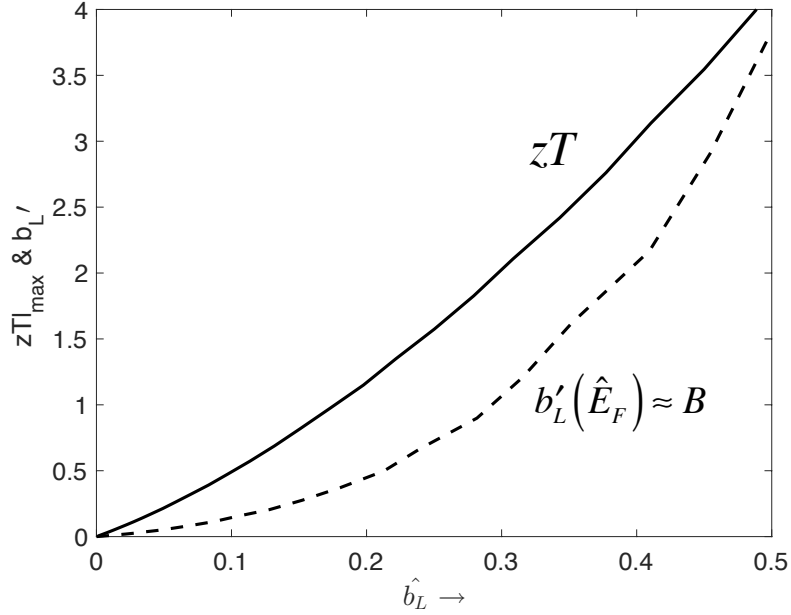

**Figure S5.** This plot shows two quantities $zT(\hat{E}_F)$ and $b'_L(\hat{E}_F)$, versus $b_L(\hat{E}_F)$ for visual comparison using a parabolic band with $r = 0$.

The figure above is a plot of $zT(\hat{E}_F)$ vs. $b_L(\hat{E}_F)$ for parabolic bands with $r = 0$ (the same as the solid line in Fig. 2 in the paper). Shown in the same figure is a plot of $B$ vs. $b_L(\hat{E}_F)$ with the same scattering, r = 0 assumed. For a given $zT$, one can read off the required $B$- or $b_L$-factor. For a given B, one can also read off the corresponding $b_L$ for an appropriate comparison. A value of $b'_L(\hat{E}_F) \approx B = 0.4$ which is a reference value often used, corresponds to $b_L(\hat{E}_F) = 0.18$ in our work. To achieve a $zT(\hat{E}_F) \approx 2$ requires a $b'_L(\hat{E}_F) \approx B \approx 1$ and a $b_L(\hat{E}_F) \approx 0.3$.